\documentclass[pdflatex,sn-nature]{sn-jnl}

\usepackage{graphicx}%
\usepackage{multirow}%
\usepackage{amsmath,amssymb,amsfonts}%
\usepackage{amsthm}%
\usepackage{mathrsfs}%
\usepackage[title]{appendix}%
\usepackage{xcolor}%
\usepackage{textcomp}%
\usepackage{manyfoot}%
\usepackage{booktabs}%
\usepackage{algorithm}%
\usepackage{algorithmicx}%
\usepackage{algpseudocode}%
\usepackage{listings}%

\makeatletter
\@twosidefalse
\@mparswitchfalse
\makeatother
\usepackage{geometry}
\usepackage{array} 
\newcolumntype{C}[1]{>{\centering\arraybackslash}p{#1}} 

\usepackage{booktabs}
\usepackage[table]{xcolor}
\usepackage{xcolor}

\theoremstyle{thmstyleone}%
\theoremstyle{thmstyletwo}%

\theoremstyle{thmstylethree}%

\begin{document}

\title[Article Title]{\vspace{-50pt} Elemental Germanium Phase-Change Memory}

\author*[1]{\fnm{Till} \sur{Zellweger}}\email{till.zellweger@iis.ee.ethz.ch}
\author[1]{Marko Mladenović}
\author[1]{Kevin Portner}
\author[1]{Christoph Weilenmann}
\author[2]{Michael Stiefel}
\author[1]{Hanglin He}
\author[1]{Klemens Bauer}
\author[3]{Luiz Felipe Aguinsky}
\author[1]{Mathieu Luisier}
\author[1]{Alexandros Emboras}

\affil[1]{\orgdiv{Integrated Systems Laboratory}, \orgname{ETH Zurich}, \orgaddress{\city{Zurich}, \country{Switzerland}}}
\affil[2]{\orgdiv{Binnig \& Rohrer Nanotechnology Center (BRNC)}, \orgname{ETH Zurich}, \orgaddress{\city{Rüschlikon}, \country{Switzerland}}}
\affil[3]{\orgdiv{DeepNano Group, James Watt School of Engineering}, \orgname{University of Glasgow},
\orgaddress{\city{Glasgow}, \country{United Kingdom}}}

\abstract{
Phase-change memory (PCM) is a mature technology for fast, scalable, non-volatile data storage, with applications spanning embedded memory, as well as in-memory and neuromorphic computing. PCM predominantly relies on chalcogenide alloys, with Ge\textsubscript{2}Sb\textsubscript{2}Te\textsubscript{5} (GST) as the industry standard. Yet in these alloys, the individual Ge, Sb, and Te atoms redistribute upon cycling, causing stochastic operation and ultimately device failure. To address this issue, elemental antimony was proposed as a PCM material, but it exhibits a metastable amorphous state that prevents reliable data retention. Moreover, tellurium and antimony can contaminate complementary metal-oxide-semiconductor (CMOS) production lines or act as unintended dopants, restricting manufacturing of PCM to dedicated fabs. Here we introduce elemental germanium (Ge) as a CMOS-native phase-change material that overcomes these fundamental limitations. In a vertical PCM cell architecture, Ge enables sub-nanosecond crystallization (240\,ps, 40 times faster than GST), non-volatile data storage with excellent thermal stability ($>$110\,$^\circ$C for 10 years vs. $\sim$87\,$^\circ$C for GST), and a resistance drift coefficient approximately 60\,\% lower than in GST. These results establish pure Ge, a standard semiconductor, as an alternative to chalcogenide phase-change materials, achieving superior performance in key metrics and enabling phase-change memory to be fabricated in standard semiconductor facilities.}


\maketitle

\section{Introduction}\label{introduction}
The rapid growth of data-centric computing and artificial intelligence places increasing demands on memory technologies in terms of speed, energy efficiency, and density \cite{fantini2025memory}. Phase-change memory (PCM) has emerged as a strong candidate to satisfy these demands, combining nanosecond-scale switching speeds approaching those of DRAM, non-volatile multi-level data storage, scalability down to a few nanometers, and the ability to perform matrix-vector multiplications directly within memory arrays, a key feature for energy-efficient neural network inference \cite{zhang2019designing,syed2025phase,portner2025actor}. PCM encodes information in the large electrical resistance contrast between the amorphous and crystalline phases of a functional material, which can be reversibly switched by short electrical pulses \cite{wuttig2007phase}. 
This large contrast is generally associated with a metavalent bonding character, a hybrid between metallic and covalent bonding, previously termed resonant bonding and shared by all established non-volatile PCM materials \cite{wuttig2018incipient,shportko2008resonant}.\\ 

The field of PCM has historically revolved around the ternary Ge-Sb-Te diagram displayed in Fig.~\ref{fig1}a \cite{wuttig2007phase,burr2016recent,fantini2020phase}. This composition space enables systematic tuning of phase-change material properties, including crystallization speed, thermal stability, and resistance contrast. The most prominent phase-change materials are found along or near the GeTe--Sb\textsubscript{2}Te\textsubscript{3} pseudo-binary line, where moving toward Sb\textsubscript{2}Te\textsubscript{3} increases crystallization speed at the cost of thermal stability \cite{fantini2020phase,zhang2019designing}. The endpoints Sb\textsubscript{2}Te\textsubscript{3} and GeTe define the extremes of this tradeoff, with Ge\textsubscript{2}Sb\textsubscript{2}Te\textsubscript{5} (GST) as the standard balanced composition \cite{fantini2020phase}. GST has been successfully deployed in commercial products ranging from rewritable optical discs to embedded non-volatile memories \cite{fantini2020phase,wuttig2007phase,hellenbrand2024progress}. Despite their success, GST and the broader chalcogenide family suffer from several limitations, including a fundamental speed-stability tradeoff, resistance drift, and random redistribution of their constituent elements under repeated cycling \cite{burr2016recent}. The latter leads to elemental and phase segregation, ultimately causing device failure \cite{syed2025phase}. In addition, all common phase-change materials contain Sb, Te, and/or Se. These elements are generally considered CMOS contaminants, thus requiring dedicated equipment and process segregation \cite{polignano2013tellurium,stumpel1988diffusion,larson1985metallic,steen2008segregation}. The added cost and complexity have so far limited the broad adoption of PCM in commercial fabrication facilities \cite{hellenbrand2024progress}.\\

Intense efforts have been directed toward alternative materials to address these limitations, including Ge-rich or doped GST to improve thermal stability and retention \cite{bala2023recent}, superlattice PCM cells to reduce switching energy and drift \cite{simpson2011interfacial}, or Sc-doped Sb\textsubscript{2}Te\textsubscript{3} (SST) to achieve fast switching speeds without sacrificing thermal stability \cite{rao2017reducing}. Nevertheless, the enhanced stability of doped and Ge-rich GST comes at the expense of crystallization speed \cite{zhang2019designing}, superlattice approaches suffer from atomic intermixing between layers \cite{khan2022unveiling}, whereas SST and, more broadly, multi-component alloy families, remain subject to redistribution of constituent elements upon cycling \cite{syed2025phase}. A conceptually different strategy was demonstrated by employing a single element as phase-change material \cite{salinga2018monatomic,shen2021elemental}. This concept inherently eliminates elemental redistribution and enables aggressive downscaling of PCM cells, unconstrained by stoichiometry requirements \cite{salinga2018monatomic,shen2021elemental}. Elemental Sb (bottom right corner of the Ge-Sb-Te triangle in Fig.~\ref{fig1}a) represents the first demonstrated single-element PCM. In addition to the benefits of its monatomic nature, it exhibits a switching speed of 242\,ps, the fastest reported for any non-volatile PCM to date \cite{shen2023toward}. However, it is also characterized by a metastable amorphous state that leads to spontaneous crystallization at room temperature \cite{chen2023suppressing}, compromising data retention. Reliable operation can only be achieved with ultra-thin film geometries \cite{chen2023suppressing}, spatial confinement \cite{leng2021crystallization}, or at cryogenic temperatures (100 K) \cite{salinga2018monatomic}. On the other hand, elemental Te (bottom left corner of the triangle) does not provide non-volatile switching and therefore finds application as a single-element selector rather than as a PCM \cite{shen2021elemental}. Moreover, both the engineered alloys and monatomic variants remain incompatible with standard CMOS processes due to contamination risk \cite{polignano2013tellurium,stumpel1988diffusion,larson1985metallic,steen2008segregation}.\\

The germanium (Ge) corner of the triangle in Fig.~\ref{fig1}a (red star) has remained unexplored as a phase-change material for different reasons. First, unlike conventional PCM materials such as GST, with partially delocalized valence electrons, characteristic of metavalent bonding (Fig.~\ref{fig1}b), Ge features covalent bonds with valence electrons localized along the inter-atomic bonds (Fig.~\ref{fig1}c). Strong covalent bonds are generally associated with a larger barrier to atomic rearrangement, slowing crystallization by orders of magnitude relative to metavalently-bonded compounds and reducing the resistance contrast between phases \cite{persch2021potential,shportko2008resonant}. Consistent with this, increasing the Ge content in GST alloys progressively reduces the resistance contrast and slows crystallization \cite{redaelli2022material}. Finally, Ge amorphization requires reaching its melting point at $\sim$1,210\,K \cite{hassion1955melting}, far above the $\sim$890\,K of GST \cite{yamada1991rapid}, raising the question of whether these temperatures are readily attainable via Joule heating during device operation. In this work, we challenge these assumptions and introduce elemental Ge as a novel phase-change material, drawing inspiration from the single-element approach pioneered by Sb. Notably, Ge is a well-established semiconductor with proven process flows in industrial foundries, both in optics as building block of high-speed photodetectors \cite{marris2018germanium}, and in electronics as stressor for $p$-type transistor channels in advanced CMOS nodes \cite{zhang2024new}. As such, elemental Ge represents the first CMOS-native phase-change material.\\

In particular, we demonstrate functional Ge PCM cells and reveal the unique advantages of this technology: (i) non-volatile storage, (ii) ultra-fast crystallization speed of 240\,ps, matching the fastest reported PCM, together with (iii) high thermal stability, breaking the speed-stability tradeoff of conventional Ge-Sb-Te alloys, and (iv) a resistance drift approximately 60\% lower than that of GST \cite{ielmini2007physical}. To prove that the underlying switching mechanism of elemental Ge is indeed phase change, we perform in-depth structural and electronic characterizations combining transmission electron microscopy (TEM) methods with molecular dynamics (MD) and quantum transport (QT) simulations, providing atomic-scale insight into the cell functionality. Overall, this work establishes elemental Ge as a fast, thermally stable phase-change material, thus extending the design space of PCM beyond the metavalent bonding paradigm. Our realization is the first step toward ultra-scalable PCM cells directly integrable into industrial semiconductor manufacturing processes.

\section{Results}\label{results}
\subsection{Concept and Device Structure}
\begin{figure}[t]
\centering
\makebox[\textwidth][c]{\includegraphics[width=1.03\textwidth]{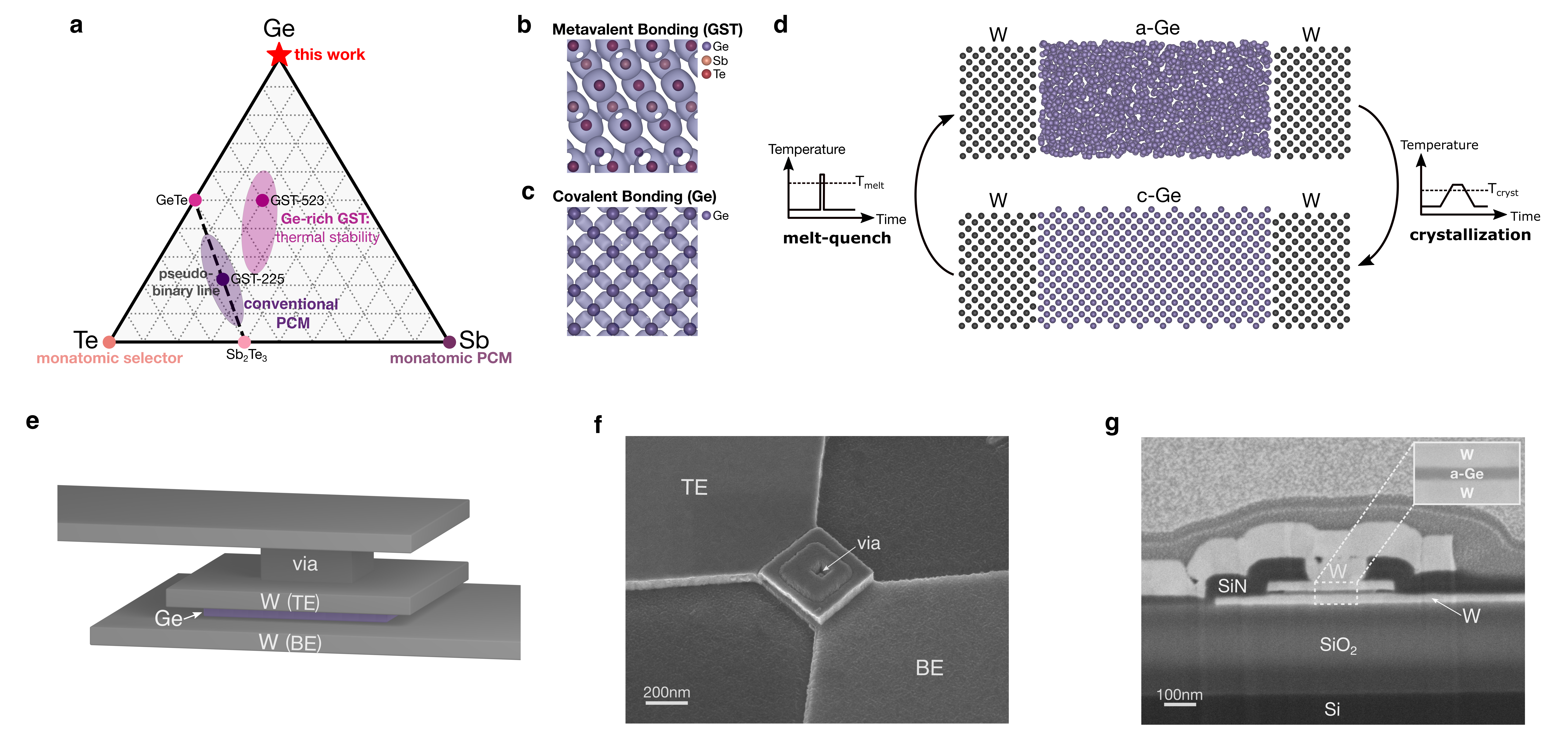}}
\caption{\textbar \textbf{ Concept and device structure of the elemental Ge phase-change memory cell.} \textbf{a,} Ternary Ge-Sb-Te composition space with prominent PCM compositions and their properties. Along the GeTe--Sb\textsubscript{2}Te\textsubscript{3} pseudo-binary line, moving toward Sb\textsubscript{2}Te\textsubscript{3} increases crystallization speed at the cost of thermal stability, while Ge\textsubscript{2}Sb\textsubscript{2}Te\textsubscript{5} (GST) represents the industry standard. Shifting the composition toward the Ge corner yields Ge-rich GST alloys (e.g., Ge\textsubscript{5}Sb\textsubscript{2}Te\textsubscript{3}\cite{yimam2022phase}) with improved thermal stability and data retention. Besides alloys, the constituent elements themselves have also been investigated. Elemental Sb represents the first single-element non-volatile PCM, while elemental Te functions as a single-element volatile selector. In this work, the Ge corner (red star) is investigated. \textbf{b,} Three-dimensional electron localization function (ELF) of crystalline GST, showing valence electrons both delocalized and partially concentrated along inter-atomic bonds, characteristic of metavalent bonding. \textbf{c,} Three-dimensional ELF of crystalline Ge, showing valence electrons strongly localized along inter-atomic bonds, characteristic of covalent bonding. \textbf{d}, Phase-change switching concept in elemental Ge, illustrated with an amorphous (a-Ge) and crystalline (c-Ge) atomic configuration of a W/Ge/W cell extracted from molecular dynamics simulations, together with the associated bias schemes. \textbf{e,} Three-dimensional rendered illustration of the Ge PCM cell design. \textbf{f,} SEM top-view image of a fabricated device, showing the top (TE) and bottom (BE) W electrodes and connecting via. \textbf{g,} FIB-SEM cross-section of the same device as in \textbf{f}, displaying the central layer stack. The inset shows a STEM image of the W/Ge/W switching region.}\label{fig1}
\end{figure}

The operating principle of our elemental Ge PCM cell (W/Ge/W structure) is illustrated in Fig.~\ref{fig1}d, showing atomic configurations in the amorphous and crystalline states obtained from atomistic simulations (see Methods). As in conventional PCM operation, a short, high-amplitude pulse with a fast trailing edge induces melt-quenching into the amorphous high-resistance state (HRS), while a lower-amplitude pulse crystallizes the material back into its low-resistance state (LRS).\\

Figure~\ref{fig1}e to g introduce the vertical Ge PCM cell design and fabricated device structure. First, a three-dimensional rendering with accurate proportions is presented in Fig.~\ref{fig1}e. Our device consists of a small Ge volume embedded between a bottom (BE) and top (TE) tungsten (W) electrode, the latter being contacted by a tungsten via. W was selected as electrode material as it is the only metal not to form inter-metallic compounds with Ge at elevated temperatures \cite{gaudet2006thin}. A scanning electron microscope (SEM) top-view image highlighting the top and bottom W electrodes and the connecting via is displayed in Fig.~\ref{fig1}f. Finally, a focused ion beam scanning electron microscope (FIB-SEM) cross-section of the full device stack is given in Fig.~\ref{fig1}g, with a scanning transmission electron microscopy (STEM) image of the central region as inset: Two 30\,nm top and bottom W electrodes enclose a 20\,nm Ge layer, initially in its amorphous phase (a-Ge). The device structure is a self-heating cell in which the nanometric Ge layer itself generates Joule heat, without a separate heater electrode as in conventional PCM cells. This geometry was chosen as it allows the full W/a-Ge/W stack to be deposited by sputtering without breaking vacuum, yielding pristine layer interfaces free of oxidation and contamination. The stack is subsequently patterned while maintaining a SiN\textsubscript{x} encapsulation throughout. The fabrication details are provided in Methods.\\

\subsection{Electrical Characterization of Elemental Ge PCM Cells}
\begin{figure}[t]
\centering
\makebox[\textwidth][c]{\includegraphics[width=1.05\textwidth]{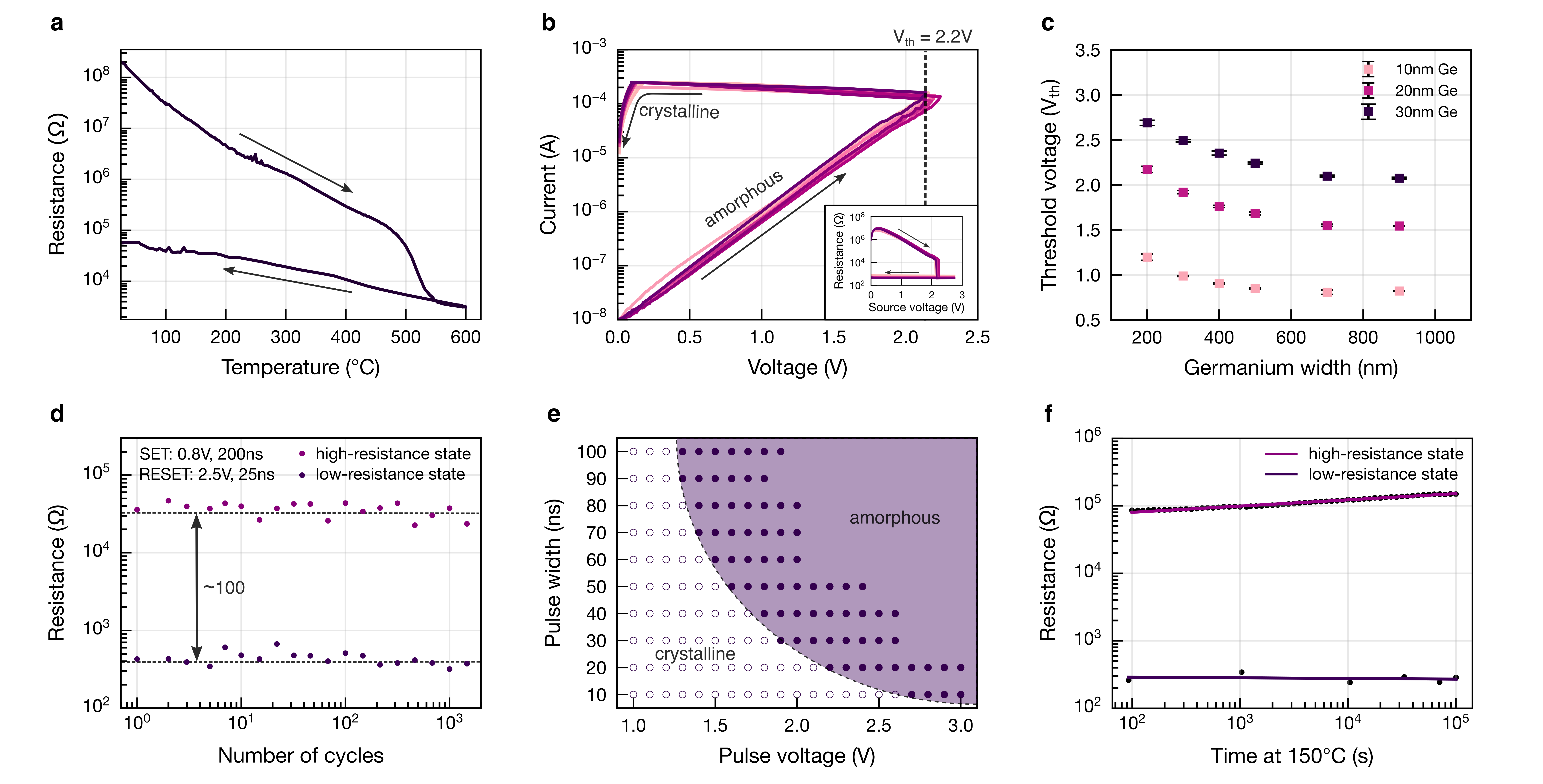}}
\caption{\textbar \textbf{ Electrical characterization of elemental Ge as phase-change material}. \textbf{a,} Resistance versus temperature of a 100\,nm sputtered Ge thin film measured at 5\,$^\circ$C\,min\textsuperscript{-1} in a nitrogen atmosphere. A resistance drop is observed around 490\,$^\circ$C, marking the amorphous-to-polycrystalline transition. The resistivities before and after the measurement are $\sim$15\,$\Omega$\,m and $\sim$3.3$\times10^{-3}$\,$\Omega$\,m, respectively. \textbf{b,} Current-voltage (I-V) characteristics of 10 pristine-state Ge PCM cells of the standard geometry ($200\times200$\,nm\textsuperscript{2}, Ge area, 20\,nm Ge thickness). Threshold switching occurs at approximately 2.2\,V with low device-to-device variability. Inset: corresponding resistance-source voltage relationship. The resistance contrast between the HRS and LRS is larger than 4 orders of magnitude. \textbf{c,} Threshold voltage $V_{\mathrm{th}}$ as a function of the Ge width and thickness with five devices per configuration, displayed as mean plus/minus one standard deviation. \textbf{d,} Reversible resistance switching over 1,500 cycles (2.5\,V, 25\,ns for the RESET and 0.8\,V, 200\,ns for the SET), as obtained after conducting the I-V SET displayed in \textbf{b}. Both resistance states remain well-separated with an on/off ratio ($R_{\mathrm{HRS}}/R_{\mathrm{LRS}}$) of approximately 100. \textbf{e,} Amorphization window for the standard device geometry as a function of the pulse amplitude and width. Filled and open circles indicate successful and unsuccessful RESET operations, respectively. The purple region indicates the conditions under which RESET is possible. \textbf{f,} Retention measurement for the HRS and LRS at 150\,$^\circ$C over $10^5$\,s. Both states remain well-separated, demonstrating thermal stability exceeding that of Ge\textsubscript{2}Sb\textsubscript{2}Te\textsubscript{5}. Extrapolation using the Ge crystallization activation energy of 3\,eV projects stable retention at $>$110\,$^\circ$C for 10 years.}\label{fig2}
\end{figure}

We characterize the electrical behavior of elemental Ge as a phase-change material in Fig.~\ref{fig2}, which includes a thin-film resistance-temperature analysis and device-level measurements (setup shown in Supplementary Fig.~S1), i.e., current-voltage, threshold voltage scaling with the Ge thickness and width, device cycling, and data retention at elevated temperatures.\\

To determine the crystallization behavior of elemental Ge, we measured the resistance of a 100\,nm sputtered Ge film, initially amorphous, as a function of temperature between 25 and 600\,$^\circ$C (Fig.~\ref{fig2}a), at a ramp rate of 5\,$^\circ$C\,min\textsuperscript{-1}, in a nitrogen atmosphere. Below the crystallization onset, the resistance decreases exponentially with increasing temperature, in line with thermally activated carrier transport in amorphous germanium \cite{walley1968electrical}. Near 490\,$^\circ$C, the resistance departs from this trend and drops abruptly, marking the amorphous-to-polycrystalline transition of Ge. This onset aligns with temperature-resolved X-ray diffraction (XRD) measurements on evaporated Ge films (Supplementary Fig.~S2). Upon cooling back to room temperature, the resistance remains low, demonstrating a non-volatile phase transition. The amorphous and polycrystalline character of the film before and after this experiment was verified by Raman spectroscopy (Supplementary Fig.~S3). Through four-point probe measurements before and after the temperature sweep, resistivities of $\sim$15\,$\Omega$\,m (amorphous) and $\sim$3.3$\times10^{-3}$\,$\Omega$\,m (polycrystalline) were found, corresponding to a $\sim$4,600$\times$ contrast. For reference, Ge\textsubscript{2}Sb\textsubscript{2}Te\textsubscript{5} exhibits around four to six orders of magnitude contrast in similar measurements \cite{redaelli2022material,friedrich2000structural,kato2005electronic}, while Ge-rich GST alloys reach one to three orders depending on the Ge concentration \cite{redaelli2022material}. Elemental Ge, with a $>$3 orders of magnitude contrast, lies between them, providing sufficient headroom for multi-level programming.\\
 
Figure~\ref{fig2}b shows the current-voltage (I-V) characteristics of 10 Ge PCM cells with a geometry referred to as standard ($200\times200$\,nm\textsuperscript{2}, Ge area, 20\,nm Ge thickness). Each curve was measured starting from the as-deposited amorphous state. All cells exhibit threshold switching at around 2.2\,V with the characteristic voltage snapback of PCM \cite{syed2025phase}, and low device-to-device variability. The inset reports the corresponding resistance-voltage (R-V) relationship extracted from the same measurement, revealing a resistance contrast from around 22\,M$\Omega$ (pristine device) to 540\,$\Omega$ (LRS) at $V_{\mathrm{read}}=0.1$\,V, which corresponds to a factor of $\sim$41,000 ($>$4 orders of magnitude). This exceeds the thin film resistivity ratio measured in Fig.~\ref{fig2}a. We attribute this larger contrast to the additional contribution of the Schottky barriers at the Ge/W electrode interfaces, which are known to present a higher barrier height in the amorphous phase of Ge than in the crystalline one \cite{oh2004metal,hull2005amorphous}. Figure~\ref{fig2}c shows the threshold voltage $V_{\mathrm{th}}$ as a function of the Ge width and thickness with five different devices per configuration. This voltage increases with the Ge layer thickness, as expected \cite{ielmini2007analytical}, and decreases with its width. The latter dependence likely reflects a statistical weakest-link effect in which a larger area enhances the probability of reaching the threshold condition.\\

We next assess the cycling behavior of the standard Ge cell geometry. Figure~\ref{fig2}d shows 1,500 RESET-SET cycles following the initial SET transition in Fig.~\ref{fig2}b. The HRS and LRS remain clearly separated throughout the measurement, with an on/off ratio ($R_{\mathrm{HRS}}/R_{\mathrm{LRS}}$) of approximately 100, demonstrating reversible switching of elemental Ge cells under unipolar pulses. Supplementary Fig.~S4 further characterizes the endurance of 10 different Ge PCM cells, cycled to switching failure at up to 5,000 cycles. This value remains below the $10^6$--$10^9$ cycles reported for optimized GST-based PCM \cite{zheng2025extended}. Scaling the contact electrode dimensions from the current 200\,nm towards the sub-50\,nm dimensions used in such high-endurance demonstrations is expected to narrow this gap.\\

Figure~\ref{fig2}e classifies the RESET operation of the standard Ge PCM cell as successful or unsuccessful depending on the amplitude and width of the applied voltage pulses. A well-defined amorphization window is observed (purple region), with shorter pulses requiring higher voltage amplitudes to achieve RESET. This dependence is the characteristic signature of Joule-heating-driven amorphization in PCM \cite{burr2016recent}. The minimum RESET energy within this characterization was measured at the shortest pulse width of 10\,ns, and amounts to $\sim$820\,pJ. It was obtained by integrating the measured transient power $P(t)$ (Supplementary Fig.~S5). For the active area of $200\times200$\,nm\textsuperscript{2}, this corresponds to an energy density of $\sim$2\,J\,cm\textsuperscript{-2}, exceeding the $\sim$1\,J\,cm\textsuperscript{-2} typically reported for GST-based PCM devices \cite{stern2021uncovering}. However, this does not represent the intrinsic limit of the Ge PCM technology. In a separate measurement, reducing the pulse width to 1\,ns led to a lowered RESET energy of $\sim$271\,pJ ($\sim$0.68\,J\,cm\textsuperscript{-2}), see 
Supplementary Fig.~S5. With improved thermal confinement, an established route to decrease the energy consumption of PCM \cite{stern2021uncovering}, even lower values could be reached.\\

Finally, the evolution of the HRS and LRS at 150\,$^\circ$C, the maximum temperature accessible with our setup, is plotted in Fig.~\ref{fig2}f over $10^5$\,seconds (approximately 1.2 days). Both states remain well-separated, demonstrating high thermal stability of the amorphous phase at elevated temperatures. For comparison, Ge\textsubscript{2}Sb\textsubscript{2}Te\textsubscript{5} fully crystallizes under these conditions \cite{choi2009effect}, and only more thermally stable alloys such as Ge-rich GST can retain a stable HRS. This indicates the potential of elemental Ge PCM cells for high-temperature embedded and automotive applications \cite{redaelli2022material}. Considering an activation energy for the crystallization of elemental Ge of approximately 3\,eV, as commonly reported in the literature \cite{chik1976annealing,germain1979crystallization}, the measured retention at 150\,$^\circ$C can be extrapolated to stable operation at $>$110\,$^\circ$C for 10 years (see Methods), surpassing the 85\,$^\circ$C 10-year benchmark of commercial non-volatile memories \cite{syed2025phase}. 

\subsection{Investigation of the Phase-Change Mechanism within Ge PCM Cells}
\begin{figure}[t]
\centering
\makebox[\textwidth][c]{\includegraphics[width=1.05\textwidth]{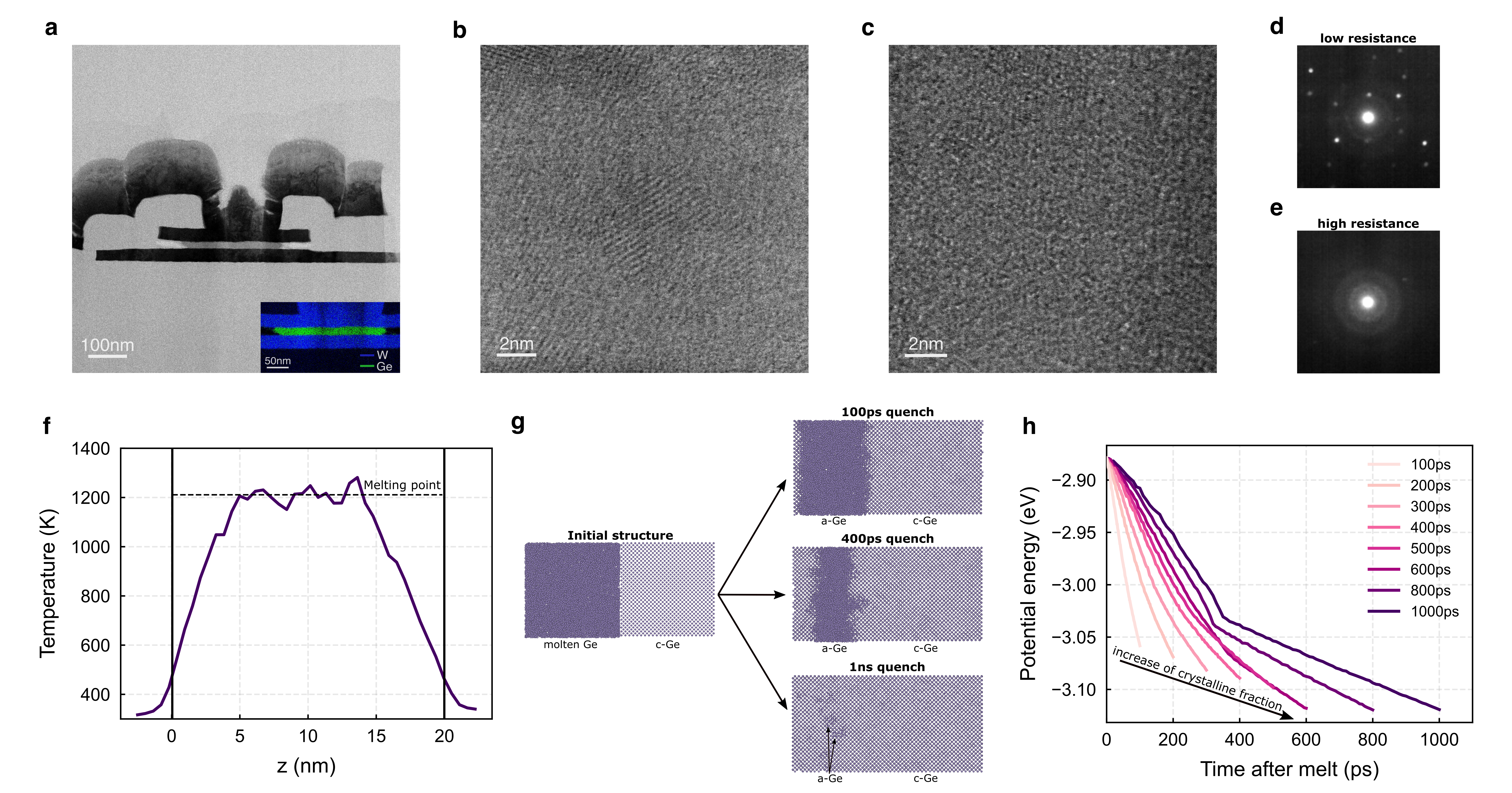}}
\caption{\textbar \textbf{ Investigation of the phase-change mechanism in elemental Ge by transmission electron microscopy (TEM) characterization and atomistic simulations}. \textbf{a}, Cross-sectional annular bright-field scanning TEM (ABF-STEM) overview of the investigated device with energy-dispersive X-ray spectroscopy (EDX) maps of germanium (green) and tungsten (blue) overlaid, confirming clear separation between the W electrodes and the Ge active layer. \textbf{b}, High-resolution ABF-STEM image of the central Ge region in the LRS, showing multiple crystalline grains identifiable by their lattice fringes. \textbf{c}, High-resolution ABF-STEM image in the HRS. The atomic arrangement is random, without discernible crystallinity. \textbf{d}, 4D-STEM electron diffraction patterns averaged over a region of $10\times10$\,nm\textsuperscript{2} for the highest-crystallinity area in the LRS, yielding sharp spots characteristic of multiple crystalline grain orientations. \textbf{e}, Corresponding diffraction pattern of the HRS, where broad diffuse rings can be seen. They are characteristic of the amorphous phase, with a weak residual spot intensity indicating a small remaining crystalline fraction. \textbf{f}, Effective lattice temperature profile across the 20\,nm crystalline Ge region between two W contacts, as obtained from coupled electron-phonon quantum transport simulation under RESET pulse conditions (amplitude of 3 V). The melting point of Ge is indicated by a dashed line. \textbf{g}, Molecular dynamics (MD) simulation of the melt-quench process using the Stillinger-Weber potential (SW) \cite{jian1990modification}. An amorphous Ge slab in contact with a crystalline Ge seed of equal volume is initialized at 1,250\,K (initial structure) and cooled to room temperature at three quench times (100\,ps, 400\,ps, and 1\,ns). The total simulation cell is $6.8 \times 6.8 \times 13.5$\,nm\textsuperscript{3}. \textbf{h}, Potential energy per Ge atom as a function of time after the melt-and-quench experiment, extracted from the same MD simulations as before. The black arrow marks the trend of increasing crystallinity (i.e., lower final $E_{\mathrm{pot}}$) with slower quench times.}\label{fig3}
\end{figure}

So far, we have established that elemental Ge devices exhibit characteristics consistent with phase-change operation, including threshold switching, reversible unipolar switching, and a well-defined amorphization window. Furthermore, resistance versus temperature measurements on Ge thin films, together with Raman spectroscopy and XRD, revealed two well-separated resistance states corresponding to crystalline and amorphous Ge. However, these measurements alone do not unambiguously confirm phase change as the origin of resistive switching in Ge cells. To exclude alternative switching mechanisms, e.g., electrochemical metallization or valence change effects \cite{ielmini2015resistive}, we present here a two-part investigation. First, direct structural evidence is obtained from cross-sectional scanning transmission electron microscopy (STEM), electron diffraction based on 4D-STEM, and energy-dispersive X-ray spectroscopy (EDX) analyses of two Ge cells in their LRS and HRS, as depicted in Figs.~\ref{fig3}a-e. In a second step, these experimental characterizations are complemented by quantum transport and molecular dynamics simulations (Figs.~\ref{fig3}f-h), which provide insight into the lattice temperature and atomic structure evolution of the active Ge volume during the RESET operation.\\

Figure~\ref{fig3}a presents a cross-sectional STEM overview of the examined device in its LRS. An EDX elemental map of its structure is shown as an inset, confirming a clear separation between the W electrode and Ge active layer, without any interdiffusion. High-resolution STEM images of the central Ge region are reported in Fig.~\ref{fig3}b for the LRS and in Fig.~\ref{fig3}c for the HRS (different device, but same geometry). The LRS is characterized by a large number of crystalline grains, clearly identifiable by their lattice fringes. In contrast, the HRS does not exhibit any such ordered arrangement. The atomic configuration appears random, without discernible crystallinity. This distinction was independently confirmed by an electron diffraction analysis using 4D-STEM, which provides spatially resolved diffraction patterns across the device cross-section. Figures~\ref{fig3}d and e display two diffraction patterns averaged over a region of $10\times10$\,nm\textsuperscript{2} corresponding to the highest-crystallinity area observed in each cell. The LRS yields sharp diffraction spots, which are indicative of multiple crystalline grain orientations, while the HRS produces broad diffuse rings that are specific to the amorphous phase. A weak residual crystalline spot is additionally visible in the HRS pattern, suggesting that a small crystalline fraction persists after the RESET pulse. Altogether, the STEM and 4D-STEM results establish that the LRS and HRS of the PCM cell correspond to predominantly crystalline and amorphous Ge, respectively, providing direct structural evidence of the underlying phase-change mechanism.\\

The melting point of Ge ($\sim$1,210\,K \cite{hassion1955melting}) substantially exceeds that of GST ($\sim$890\,K \cite{yamada1991rapid}), raising the question of whether Joule heating in a 20\,nm-thick Ge layer is sufficient to reach this high temperature during device operation. To address this question, we performed coupled electron-phonon quantum transport simulations (see Methods) that take the applied voltage under RESET conditions as input and return the electronic current $I_{\mathrm{d}}$ as well as the effective lattice temperature $T_{\mathrm{eff}}$ of Ge. The simulated structure is described in Supplementary Fig.~S6. At a pulse amplitude of 3\,V, $I_{\mathrm{d}} \approx 30$\,mA flows through the standard Ge cell geometry, in agreement with experimental data (Supplementary Fig.~S5). At the same time, the spatially-resolved $T_{\mathrm{eff}}$, shown in Fig.~\ref{fig3}f, exceeds the melting point of Ge in the center of the cell, confirming that melting of Ge can be achieved under the RESET bias conditions.\\

As a final validation step, after determining that elemental Ge can be melted during cell operation, we tested the feasibility of stabilizing the amorphous phase via melt-quench MD simulations using the Stillinger-Weber inter-atomic potential \cite{jian1990modification} (see Methods). To mimic melt-quench conditions in a device where residual crystalline Ge acts as a recrystallization seed, an amorphous Ge slab is placed in contact with a crystalline Ge seed of equal volume, initialized at 1,250\,K and quenched at varying rates to room temperature. This procedure melts the amorphous slab while preserving the crystalline seed \cite{tipeev2020crystal}. Figure~\ref{fig3}g reports the initial atomic structure at 1,250\,K along with the resulting ones at room temperature for three quench times. For a 100\,ps quench, the molten Ge volume almost entirely stabilizes into amorphous Ge, while for a 1\,ns quench, recrystallization from the seed dominates and only a negligible amorphous fraction remains at room temperature. Figure~\ref{fig3}h shows the calculated potential energy per atom $E_{\mathrm{pot}}$ over time for a wider range of quench times. Since amorphous Ge has a higher $E_{\mathrm{pot}}$ than its crystalline counterpart, the evolution of $E_{\mathrm{pot}}$ over time serves as a proxy for crystallinity. The simulations reveal that slower quenches yield lower final $E_{\mathrm{pot}}$ at room temperature (i.e., higher crystallinity after melt-quench), as indicated by the black arrow and consistent with the snapshots in Fig.~\ref{fig3}g. These findings are supported by similar results obtained from MD simulations based on a first-principles machine-learned inter-atomic potential \cite{jin2026data} (Supplementary Fig.~S7). The simulations thus attest that the amorphous phase of elemental Ge can be stabilized via melt-quench processes.

\subsection{Benchmarking Elemental Ge Against Established Phase-Change Materials}
\begin{figure}[t]
\centering
\makebox[\textwidth][c]{\includegraphics[width=1.05\textwidth]{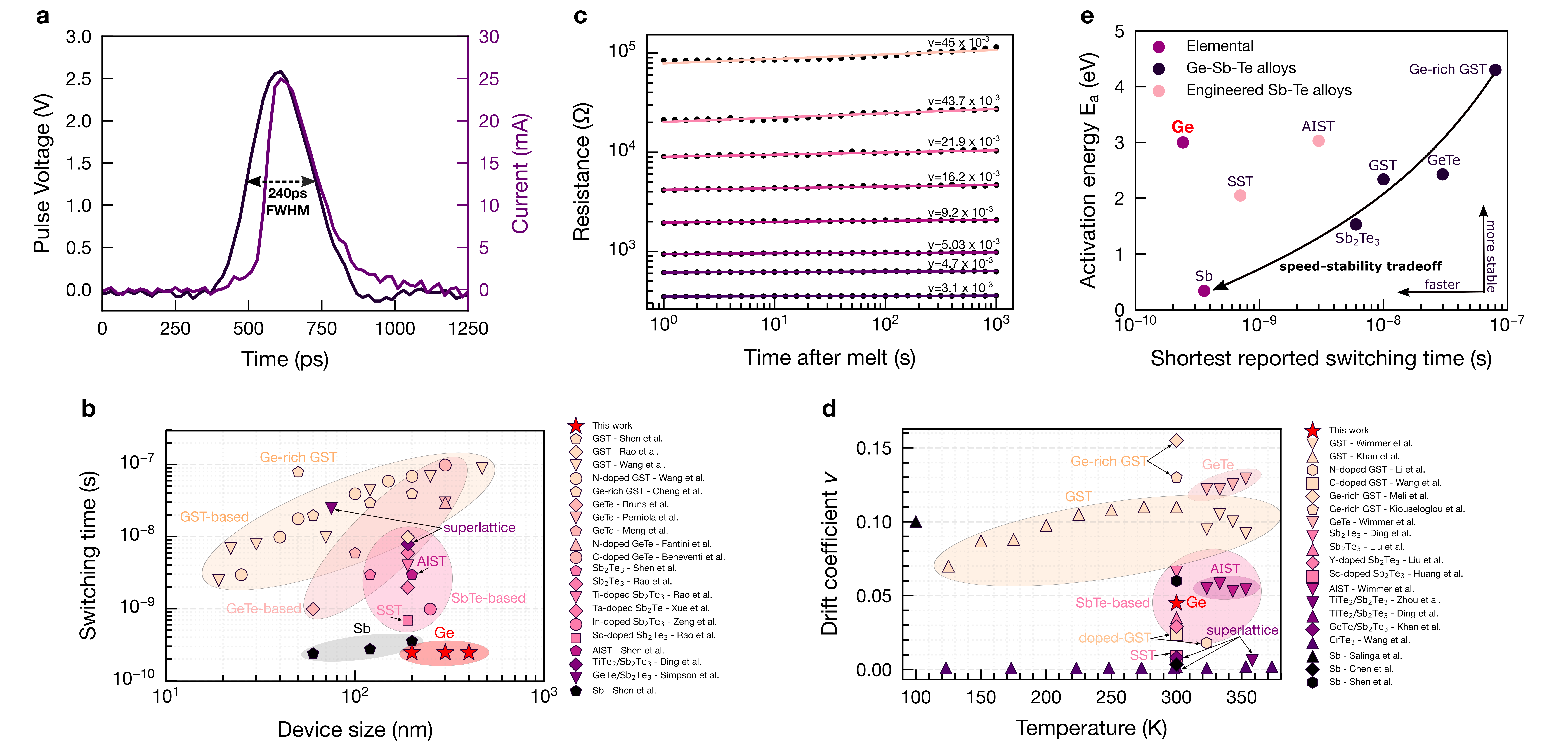}}
\caption{\textbar \textbf{ Benchmarking of elemental Ge PCM cells against established phase-change materials}. \textbf{a,} High-speed crystallization of a standard Ge PCM cell. Applied voltage pulse (black) and measured current response (dark purple) at 240\,ps FWHM. \textbf{b,} Switching time, defined as minimum pulse width to achieve crystallization, as a function of the device size for seven phase-change material classes: GST \cite{shen2023toward,rao2017reducing,wang2008fast,wang2012enabling,cheng2011high}, GeTe \cite{bruns2009nanosecond,perniola2010electrical,meng2023gete,fantini2010n,beneventi2011carbon}, Sb-Te \cite{shen2023toward,rao2017reducing,xue2021ta,zeng2024fast}, Ag-In-Sb-Te (AIST) \cite{shen2023toward}, superlattice-like PCM \cite{ding2019phase,simpson2011interfacial}, elemental Sb \cite{shen2023toward}, and Ge (this work). Material classes are enclosed by colored ovals, the subclasses superlattice-like PCM, SST and AIST are indicated by labeled arrows. The device size corresponds to the bottom electrode diameter for mushroom-type cells and to the lateral width of the Ge active region for the Ge PCM cells reported here. In both cases this dimension defines the thermally active volume governing crystallization speed. \textbf{c,} Resistance as a function of time for the Ge PCM cell programmed to a series of different states by single RESET pulses of amplitudes 2--2.35\,V, each applied from the LRS. All curves are fitted according to the power-law relation $R(t)=R_0(t/t_0)^\nu$. The extracted drift coefficients $\nu$ are indicated in each case. \textbf{d,} Reported drift coefficients as a function of measurement temperature for the same seven material classes \cite{wimmer2014role,khan2020resistance,li2018understanding,wang2022reliable,meli2023,Kiouseloglou2014,ding2019phase,liu2021multi,huang2023nanoscale,zhou2020resistance,khan2021ultralow,salinga2018monatomic,shen2023toward,chen2023suppressing} and CrTe\textsubscript{3} \cite{wang2025amorphous}, compared to elemental Ge. \textbf{e,} Crystallization activation energy $E_{\mathrm{a}}$ \cite{chik1976annealing,choi2009effect,hashimoto1980effect,zuliani2013overcoming,whab2025elucidating,wang2008temperature,njoroge2004influence,chen2022anomalous} versus shortest reported crystallization time at 200\,nm device size \cite{rao2017reducing,shen2023toward,cheng2011high,perniola2010electrical} (as defined above) for each PCM material. The $E_{\mathrm{a}}$ values are from resistance-based thin-film measurements in the low-temperature or low-heating-rate regime to reflect retention-relevant crystallization kinetics, except for Sb, which was analyzed in situ during deposition. For Ge-rich GST and GeTe, the nearest available device size is used (50\,nm and 300\,nm, respectively) as no 200\,nm data is available. The black arrow indicates the speed-stability tradeoff governing Ge-Sb-Te alloys and elemental Sb.}\label{fig4}
\end{figure}
Having established the electrical operation of our elemental Ge PCM cell and confirmed phase change as the origin of resistive switching, we now assess the applicability of elemental Ge as a new phase-change material by benchmarking its crystallization speed, resistance drift, and thermal stability against the state of the art.\\

To determine the crystallization speed, we characterized devices with Ge widths of 200, 300, and 400\,nm in the pristine, as-deposited amorphous state. This can be seen as a lower bound on the maximum speed, as already melt-quenched amorphous states are known to crystallize substantially faster \cite{coombs1995laser}. The shortest pulse achievable by our arbitrary waveform generator (AWG), 240\,ps full width at half maximum (FWHM), leads to crystallization for all device sizes considered, as shown in Fig.~\ref{fig4}a for the standard geometry (200\,nm Ge width). Analogous current transients measured at different pulse amplitudes for each device size are shown in Supplementary Fig.~S8, and the corresponding post-pulse resistance as a function of pulse amplitude in Supplementary Fig.~S9. In Fig.~\ref{fig4}a, no discernible delay is observed between the voltage pulse reaching $V_{\mathrm{th}}$ and the rise of the measured current through the Ge cell, indicating that crystallization occurs with negligible incubation time. The sub-nanosecond crystallization speed of elemental Ge is consistent with an explosive crystallization mechanism, which has been well documented in amorphous elemental Ge thin films \cite{egan2019novel}: latent heat released at the advancing crystallization front superheats the surrounding amorphous material and drives self-propagating crystal growth. Additionally, Ge exhibits a temperature-dependent activation energy with $\sim$1.4\,eV at higher temperatures ($\gtrsim$360\,$^\circ$C) \cite{germain1979crystallization} relevant to crystallization. It is substantially lower than the previously mentioned $\sim$3\,eV relevant to data retention, and may further contribute to the fast switching kinetics of Ge. Explosive crystallization has also been reported for elemental Sb \cite{bostanjoglo1981impulse}, though the ultra-fast switching observed in these devices is generally attributed to a growth-dominated crystallization mechanism \cite{shen2023toward}.\\

Figure~\ref{fig4}b compares the required pulse width for crystallization of elemental Ge PCM cells against the state of the art of various other phase-change materials, as a function of the device size. The plot highlights seven material classes: GST \cite{rao2017reducing,shen2023toward,wang2008fast,wang2012enabling,cheng2011high}, GeTe \cite{bruns2009nanosecond,perniola2010electrical,meng2023gete,fantini2010n,beneventi2011carbon}, Sb-Te \cite{shen2023toward,rao2017reducing,xue2021ta,zeng2024fast}, Ag-In-Sb-Te (AIST) \cite{shen2023toward}, superlattice-like PCM \cite{ding2019phase,simpson2011interfacial}, elemental Sb \cite{shen2023toward}, and Ge. A clear trend of faster crystallization at smaller device dimensions can be identified. It is particularly evident for GST and GeTe, with GeTe possessing an even steeper size dependence than GST, as its growth-dominated crystallization mechanism more strongly scales with device dimensions than the nucleation-dominated crystallization of GST \cite{bruns2009nanosecond}. Elemental Ge achieves 240\,ps across all three device sizes investigated (200, 300, and 400\,nm). At an equivalent device size of 200\,nm, this is faster than the 359\,ps reported for elemental Sb \cite{shen2023toward}. At 240\,ps, elemental Ge matches the current record of 242\,ps set by elemental Sb at 60\,nm \cite{shen2023toward}, i.e., in device sizes around 3 to 7 times smaller than ours. It should be reemphasized that the measured 240\,ps is instrument-limited and shows no size dependence, indicating the true crystallization speed of elemental Ge lies below 240\,ps.\\

Next, we investigate the state-dependent resistance drift of elemental Ge PCM cells at room temperature. For each measured state, a single RESET pulse was applied from the LRS, with an amplitude varied between 2 and 2.35\,V to program progressively higher initial resistances $R_0$. Figure~\ref{fig4}c shows the resistance as a function of time over $10^3$\,s for each programmed state. The resulting data points were fitted with the well-established drift equation $R(t)=R_0(t/t_0)^\nu$ \cite{pirovano2004low}. The drift coefficient $\nu$ is indicated alongside the corresponding trace. It rises from negligible values at low resistance states ($\sim$3$\times$10\textsuperscript{-3}), consistent with a predominantly crystalline active volume, to $\nu \approx 0.045$ in the HRS at $R_0 \approx 90$\,k$\Omega$. This HRS drift coefficient is 59\% lower than typically reported for Ge\textsubscript{2}Sb\textsubscript{2}Te\textsubscript{5} ($\nu \approx 0.11$) \cite{ielmini2007physical,wimmer2014role,khan2020resistance}.\\

Figure~\ref{fig4}d places this HRS drift coefficient in the context of established phase-change materials, where the $\nu$ values are plotted as a function of measurement temperature for the same material classes as in Fig.~\ref{fig4}b, with the addition of the recently demonstrated CrTe\textsubscript{3} \cite{wang2025amorphous}. Three tiers of drift magnitudes can be seen. At the upper end, Ge-rich GST \cite{meli2023,Kiouseloglou2014} and GeTe \cite{wimmer2014role} have the highest drift coefficients, reaching up to $\nu > 0.15$, while Ge\textsubscript{2}Sb\textsubscript{2}Te\textsubscript{5} \cite{wimmer2014role, khan2020resistance} spans a range between 0.07 and 0.11, depending on the temperature, with a clear increase at elevated temperatures, in line with a thermally-accelerated structural relaxation of the amorphous network \cite{khan2020resistance}. The next tier is characterized by $\nu$ values $\sim$0.02--0.07 around room temperature, well below GeTe and Ge\textsubscript{2}Sb\textsubscript{2}Te\textsubscript{5}. It includes doped GST \cite{li2018understanding,wang2022reliable}, Sb-Te-based alloys \cite{ding2019phase,liu2021multi, huang2023nanoscale}, and AIST \cite{wimmer2014role}. Elemental Ge, at $\nu \approx 0.045$, falls within this tier without any compositional engineering or geometric confinement. Elemental Sb shows $\nu \approx$ 0.06--0.1 with conventional geometries \cite{salinga2018monatomic,shen2023toward}, reduced to 0.0034 in ultra-thin (4\,nm) films with SiO\textsubscript{2} interfaces \cite{chen2023suppressing}, where interfacial effects dominate and suppress structural relaxation. The lowest reported $\nu$ belong to superlattice-like PCMs \cite{zhou2020resistance,ding2019phase,khan2021ultralow}, SST \cite{huang2023nanoscale}, and CrTe\textsubscript{3} \cite{wang2025amorphous}, with essentially drift-free operation.\\

The crystallization activation energies $E_{\mathrm{a}}$ of the same material classes as before are compared in Fig.~\ref{fig4}e. This quantity is plotted against the shortest required pulse width for crystallization in device demonstrations, thus relating thermal stability and crystallization speed. Since $E_{\mathrm{a}}$ strongly depends on the measurement technique and film geometry, we restricted the comparison to resistance-based measurements of uncapped thin films ($>$10\,nm) employing an Arrhenius or Kissinger analysis. Values from the low-temperature or low-heating-rate regime were selected to best reflect retention-relevant crystallization kinetics. For elemental Sb, which spontaneously crystallizes at room temperature, $E_{\mathrm{a}}$ was instead determined from an in-situ optical analysis in the deposition chamber \cite{hashimoto1980effect}. Crystallization speeds are extracted from devices with $\sim$200\,nm width, matching the dimension of our elemental Ge cells and the range for which the most literature data is available. Figure~\ref{fig4}e clearly highlights the speed-stability tradeoff governing elemental Sb and all Ge-Sb-Te alloys in the ternary diagram, indicated by the black arrow. In the top right corner, Ge-rich GST represents the most thermally stable alloy with the slowest reported crystallization speed, while elemental Sb occupies the bottom left with the fastest reported crystallization speed but lowest thermal stability. A few carefully engineered phase-change materials such as AIST and SST do not follow this trend. They both exhibit a pronounced fragile-to-strong (FTS) crossover that decouples low-temperature stability from high-temperature crystallization kinetics \cite{chen2019kinetics,orava2015fragile}. In SST, Sc-stabilized structural precursors additionally shorten nucleation time \cite{chen2019kinetics}. Elemental Ge also breaks the speed-stability tradeoff, occupying a position distinct from all other materials in Fig.~\ref{fig4}e: it combines the fastest reported crystallization speed at 200\,nm device size with a high retention-relevant $E_{\mathrm{a}}$ of $\sim$3\,eV. This may be attributed to explosive crystallization and the temperature-dependent $E_{\mathrm{a}}$ of Ge ($\sim$1.4\,eV at high temperatures versus $\sim$3\,eV at low temperatures \cite{germain1979crystallization}), with the latter potentially giving rise to non-Arrhenius crystallization kinetics analogous to the FTS crossover of SST and AIST.

\section{Conclusion}\label{conclusion}
In summary, this work establishes elemental Ge, a well-known semiconductor in industrial CMOS processes, as a functional phase-change material, with melt-quench amorphization and recrystallization confirmed by cross-sectional STEM, 4D-STEM electron diffraction, quantum transport and MD simulations. The Ge PCM cells break the speed-stability tradeoff that constrains Ge-Sb-Te alloys, simultaneously achieving a crystallization speed of $\leq$240\,ps, matching the fastest reported non-volatile PCM \cite{shen2023toward}, and excellent thermal stability with a 10-year retention temperature above 110\,$^\circ$C. Notably, the drift coefficient of 0.045 is approximately 60\% lower than in Ge\textsubscript{2}Sb\textsubscript{2}Te\textsubscript{5} \cite{ielmini2007physical}, achieved without compositional engineering or geometric confinement. Going forward, introducing a dedicated nanoheater and improved thermal confinement are expected to bring the endurance toward the $10^6$--$10^9$ cycles typical of optimized GST mushroom-type cells \cite{zheng2025extended}. Overall, these results position elemental Ge as a promising candidate for high-temperature embedded memory, high-speed non-volatile storage, or analog in-memory computing, fully integrable with CMOS.

\section{Methods}\label{methods}

\subsection{Device Fabrication}
The Ge PCM cells were fabricated on silicon substrates with 300\,nm thermal oxide. Stress-minimized W/a-Ge/W stacks (30\,nm/$t_{\mathrm{Ge}}$/30\,nm) with $t_{\mathrm{Ge}} \in \{10, 20, 30\}$\,nm were deposited by magnetron sputtering (DC for W, RF for a-Ge) in a multi-chamber system without breaking vacuum, ensuring clean interfaces between the Ge active layer and the W electrodes and preventing oxygen incorporation. The top W and the Ge layers were patterned by reactive-ion etching (RIE) in a single step to define isolated rectangular mesa structures on top of the bottom 30\,nm W layer. The etch recipe removes Ge more rapidly than W, creating a lateral underetching of the Ge layer and a further reduced active Ge volume confined between the two W electrodes. The mesas were capped with 50\,nm of low-stress SiN\textsubscript{x} deposited by plasma-enhanced chemical vapor deposition (PECVD) at 300\,$^\circ$C to passivate the active area. The SiN\textsubscript{x} and bottom W layer were subsequently patterned by separate RIE steps to define the bottom electrode, followed by a second 50\,nm low-stress SiN\textsubscript{x} PECVD deposition for electrical isolation of the bottom electrode from the subsequently deposited top metal layer. Vias were opened through the SiN\textsubscript{x} by RIE to expose the top W of the mesa and the bottom electrode contact area. Finally, 130\,nm of W were sputter-deposited to fill the vias and form the contact pads, and patterned by RIE to define contacts for the top and bottom electrodes in a radio-frequency (RF) compatible coplanar waveguide (CPW) geometry.

\subsection{Electrical Device Characterization}
The current-voltage (Figs.~\ref{fig2}b,c) and the retention measurements at 150\,$^\circ$C were performed using a commercial probe station (Cascade Microtech Summit 9600 thermal probe station) with a temperature controller (Temptronic TPO3010B) connected to a precision source measure unit (Keysight M9601A), shown in Supplementary Fig.~S1a.\\

All pulsed electrical measurements, including device cycling (Fig.~\ref{fig2}d), amorphization window characterization (Fig.~\ref{fig2}e), and resistance drift (Fig.~\ref{fig4}c), were conducted with an RF setup shown in Supplementary Fig.~S1b. An AWG (Active Technologies AWG-5064) was used to apply voltage pulses, while a precision source measure unit (Keysight B2902A) was utilized to measure the device resistance before and after each pulse, with the active input selected via an RF multiplexer (Keysight DAQM905A). Signals were routed to the device through 40\,GHz probes (GGB Industries Picoprobe model 40A) landing on the coplanar waveguide pads. The output signal was recorded directly by an oscilloscope (Rohde \& Schwarz RTE1104). A second AWG channel carrying the same signal was connected to the oscilloscope as a voltage reference.\\

For the crystallization speed measurements (Fig.~\ref{fig4}a), the AWG was connected directly to the device via 40\,GHz probes, with the output recorded by a 20\,GHz oscilloscope (Tektronix DPO72004). The device resistance before and after each pulse was measured separately with a precision source measure unit (Keysight B2902A).

\subsection{Device Retention Measurements}
Retention measurements were performed on two devices programmed to the HRS and LRS, respectively, using the commercial probe station described above. Both devices were measured in the same bake at $150\,^\circ$C: the HRS device resistance was recorded at logarithmically spaced time intervals over 10\textsuperscript{5}\,s, with the LRS device resistance measured at regular intervals in between HRS measurements to confirm the stability of the SET state. The 10-year retention temperature was extrapolated using an Arrhenius model for the crystallization time:
\begin{equation}
    t_{\mathrm{fail}}(T) = \tau_0 \exp\!\left(\frac{E_\mathrm{a}}{k_\mathrm{B} T}\right),
    \label{eq:retention}
\end{equation}
where $\tau_0$ is a pre-exponential factor, $E_{\mathrm{a}} \approx 3$\,eV is the crystallization activation energy, $k_\mathrm{B}$ is the Boltzmann constant, and $T$ is the absolute temperature. Since no crystallization was observed over the full 10\textsuperscript{5}\,s measurement window at $150\,^\circ$C, the pre-exponential factor $\tau_0$ was obtained by setting $t_{\mathrm{fail}} = 10^5$\,s and $T = 150\,^\circ$C in Eq.~\ref{eq:retention}. The 10-year retention temperature was then obtained as a lower bound by solving Eq.~\ref{eq:retention} for $T$ with $t_{\mathrm{fail}} = 10$\,years.

\subsection{Resistance versus Temperature Measurements}
Resistance versus temperature measurements of the Ge thin films were executed in a nitrogen glovebox via a precision hotplate (Harry Gestigkeit PZ28-3TD with PR3-3T controller) and a source measure unit (Keithley 2400). Current reversal at each data point canceled thermoelectric offsets arising from the Seebeck effect, and the source current was adjusted according to the resistance range to maintain resistance measurement accuracy across the full dynamic range of the phase transition. The thin film resistivity before and after the temperature ramp was determined separately using a Jandel Engineering collinear four-point probe setup.

\subsection{TEM Characterization}
Cross-sectional TEM lamellae of devices in the high- and low-resistance state were prepared separately by focused ion beam (FIB) milling using a TFS Helios 600i. Annular bright-field STEM overview images, high-resolution STEM images, and energy-dispersive X-ray spectroscopy (EDX) maps were acquired on a JEOL JEM-ARM300F Grand ARM operating at 300\,kV, equipped with a cold field emission gun and a spherical aberration corrector, enabling atomic-resolution imaging. Four-dimensional STEM (4D-STEM) electron diffraction patterns were acquired on a JEOL JEM-F200.

\subsection{Electronic Structure Calculations}
To demonstrate the differences in bonding characteristics between the crystalline Ge and the crystalline Ge\textsubscript{2}Sb\textsubscript{2}Te\textsubscript{5} (Fig.~\ref{fig1}b and c), the electronic localization functions (ELF) of both materials were calculated using the Density Functional Theory (DFT) approach implemented in the CP2K package \cite{cp2k}. Prior to the ELF computations, both atomic structures were fully relaxed within the same DFT framework. For both calculation types, the Perdew–Burke–Ernzerhof functional (PBE) \cite{pbe} exchange–correlation functional and the DZVP basis set \cite{dzvp} were employed. The plane-wave cutoffs used for mapping the Gaussian-type orbitals onto the grid and for the charge density were set to 50 Ry and 300 Ry, respectively. The same setup was used to generate the W/a-Ge/W and W/c-Ge/W device structures (Fig.~\ref{fig1}d), which was accomplished by optimizing the distance between the contacts and Ge and assembling the full stack. The c-Ge structure was obtained by taking the unit cell from the Materials Project \cite{materialsproject} database and constructing a $10 \times 5 \times 5$ supercell. The same supercell was then subjected to the melt-quench protocol, described at the end of the Molecular Dynamics Simulations subsection, to obtain the a-Ge structure.

\subsection{Quantum Transport Simulations}
To determine the magnitude of the self-heating effect in W/Ge/W cells under RESET conditions, coupled electro-thermal simulations were conducted with a full-band and atomistic quantum transport solver \cite{rhyner2014}. This tool relies on an $sp^3d^5s^*$ tight-binding (TB) model without spin-orbit coupling to describe the electronic properties of Ge \cite{boykin2004} and on a valence-force-field (VFF) approach including four types of atomic interactions to model the phonon characteristics of this material \cite{sui1993}. The resulting Hamiltonian and dynamical matrices serve as inputs to electron ($G(E)$) and phonon ($D(\omega)$) non-equilibrium Green's functions (NEGF), respectively, which are coupled with each other through scattering self-energies, $\Sigma(E)$ for electrons and $\Pi(\omega)$ for phonons. By self-consistently solving the coupled $(G,\Sigma,D,\Pi)$ system of equations for all possible energies $E$ and frequencies $\omega$, the electronic current ($I_{\mathrm{d}}$), the energy current ($I_{\mathrm{dE}}$), the electron/hole densities ($n/p$), and the phonon population ($\rho_{\mathrm{ph}}$) of the structure of interest are obtained. Importantly, through the externally applied voltage (3 V in our case) and the electron-phonon coupling, both $n/p$ and $\rho_{\mathrm{ph}}$ are driven out of equilibrium and energy exchanges occur between them. For example, through phonon emission, electrons injected from the grounded left side of the W/Ge/W cell lose energy when moving towards the right side of this device to which a voltage of 3\,V is applied (Supplementary Fig.~S6a). As a consequence, the local phonon population increases. This quantity can be converted into a spatially-resolved effective lattice temperature (Fig.~\ref{fig3}f). Overall, our model ensures both electronic and energy current conservation, two critical conditions (Supplementary Fig.~S6b). Additional simulation results and details are provided in Supplementary Fig.~S6, while further details on the simulation approach are given in \cite{rhyner2016}.

\subsection{Molecular Dynamics Simulations}
To assess whether amorphous Ge can be stabilized at device-relevant cooling rates (Fig.~\ref{fig3}g and h), melt-quench simulations were performed using two inter-atomic potentials to verify robustness: the empirical Stillinger-Weber potential (SW) \cite{jian1990modification} and the machine-learned Gaussian approximation potential (GAP) \cite{jin2026data}. For the SW simulations presented in the main text, the a-Ge structure was created using isotropic NPT dynamics by melting the crystalline structure at a temperature above the melting point (5,000\,K), followed by cooling to 300\,K at a rate of 4.7\,K\,ps\textsuperscript{-1}, a subsequent equilibration at 300\,K for 50\,ps, and an energy minimization procedure. The amorphous structure was then placed in direct contact with the crystalline structure of the same size. For the melt-quench simulation, the resulting bilayer system ($6.8 \times 6.8 \times 13.5$\,nm\textsuperscript{3}, 27{,}648 atoms) was first equilibrated at a temperature close to the melting point (1,250\,K) for 10\,ns, followed by quenching to 300\,K over time intervals ranging from 100\,ps to 1\,ns. At each timestep, the potential energy $E_{\mathrm{pot}}$ and the kinetic energy $E_{\mathrm{kin}}$ were calculated, together constituting the total energy of the system. Whereas $E_{\mathrm{kin}}$ reflects atomic thermal motion, $E_{\mathrm{pot}}$ arises from inter-atomic interactions and therefore serves as a proxy for crystallinity. The analogous procedure using the GAP potential and the corresponding results are presented in Supplementary Fig.~S7. The GAP potential was also employed to obtain the a-Ge structure for the W/a-Ge/W device model. In this case, the c-Ge was melted at 1,500\,K for 100\,ps, followed by quenching to 300\,K over 100\,ps, and finally annealing at 300\,K for 100\,ps.

\backmatter
\bmhead{Supplementary information}
Supplementary information is available for this paper.

\bmhead{Acknowledgments}
The authors acknowledge funding from the ALMOND project supported by the SNSF Sinergia program (grant no. 198612). This work was also supported in part by the Swiss State Secretariat for Education, Research, and Innovation (SERI) through the SwissChips research project, by the Swiss National Supercomputing Center (CSCS) under project lp16, and by the Austrian Science Fund (FWF) under project J-4686. Some results were obtained using the ARCHIE-WeSt High Performance Computer based at the University of Strathclyde. The work was carried out at the Binnig and Rohrer Nanotechnology Center (BRNC), at the Scientific Center for Optical and Electron Microscopy (ScopeM), and at ETH Zurich. We thank the Cleanroom Operations Team of the BRNC for fabrication support, and Christian Zaubitzer and Milivoj Plodinec at ScopeM for lamella preparation and TEM operation, respectively.

\bmhead{Author contributions}
T.Z. conceived the project, designed and fabricated the devices, and performed the experiments. M.M. prepared the valence ELF and atomistic device structures. M.M. and L.F.A. performed the MD melt-quench simulations. L.F.A. developed the bilayer MD simulation approach. M.L. ran the quantum transport simulations. K.P. helped design the device structure and with fabrication. C.W. co-developed the measurement setups. M.S. helped with fabrication and prepared the FIB-SEM image. K.P., C.W., H.H., and K.B. participated in the data analysis. All authors contributed to the editing of the manuscript. M.L. and A.E. supervised the project.

\bmhead{Competing interests}
The authors declare no competing interests.

\newpage


\bibliography{bibliography.bib}

\end{document}